\documentclass[10pt,twocolumn,showpacs,longbibliography,amsmath,amssymb,osajnl,floatfix,superscriptaddress]{revtex4-2}

\usepackage{graphicx}
\usepackage{dcolumn}
\usepackage{bm}
\usepackage{color}
\usepackage{txfonts}
\usepackage{microtype}

\usepackage{epstopdf}

\begin{document}


\title{Effects of angular spread in nonlinear Compton scattering}

\author{Ya-Nan Dai}
\affiliation{Department of Physics, Shanghai Normal University, Shanghai 200234, China}
\author{Jing-Jing Jiang}
\affiliation{Department of Physics, Shanghai Normal University, Shanghai 200234, China}
\author{Yu-Hang Jiang}
\affiliation{Department of Physics, Shanghai Normal University, Shanghai 200234, China}
\author{Rashid Shaisultanov}
\affiliation{Extreme Light Infrastructure ERIC, Za Radnici 835, Dolni Brezany, 25241, Czech
Republic}
\author{Yue-Yue Chen}
\email{yue-yue.chen@shnu.edu.cn}
\affiliation{Department of Physics, Shanghai Normal University, Shanghai 200234, China}
\date {\today}

\begin{abstract}
We investigated the effects of the momentum spread of photon around incoming electrons during nonlinear Compton scattering of an elliptically polarized laser off an ultrarelativistic electron beam. It has been assumed to be a good approximation to neglect the angular spread in the strong-field QED codes considering its smallness in relativistic regime. Here, we scrutinize the validity of this approximation in the nonlinear Compton scattering. For our purpose, we improved the fully electron spin- and photon polarization- resolved Monte Carlo simulation method by
employing the angle-resolved probability for high-energy photon emission in ultrastrong laser fields. The quantum operator method introduced by Baier and Katkov is employed for 
calculation of the probability within the quasiclassical approach and the local constant field approximation. 
Our simulation shows that the angular spread at emission has notable effects on angular distribution and polarization of out-going particles.
The width of angular distributions for electrons and emitted photons are increased by $55\%$ and $22\%$, respectively. Meanwhile, the electrons polarization is reduced by $11\%$ due to the correction of radiation reaction force, while the average polarization of photons is insensitive to the angular spread at emissions.
\end{abstract}

\maketitle

\section{Introduction}\label{sec:level1}
Polarized unltrarelativistic particle beams have important applications in many fields thanks to the unique additional spin information \cite{anthony2004observation,moortgat2008polarized,abe1995precision,alexakhin2007deuteron}. For instance, spin-polarized electron (positron) beams can be used in
polarized deep-inelastic lepton-nucleon scattering to provide information on the spin structure of the nucleon, and to search for new physics beyond the standard model\cite{androic2013first,adams1997spin,kuhn2009spin,herczeg2003cp,ananthanarayan2018inclusive,godbole2006lepton}. Polarized $\gamma$ photons are excellent probe of protons, neutrons, and nuclei partially due to the smallness of photon-hadron cross section\cite{schaerf2005polarized}, and could be used for detecting QED vacuum taking advantage of the enhancement of QED vacuum nonlinearity for energetic photons\cite{brodin2001proposal,nakamiya2017probing,
bragin2017high,king2016vacuum,ataman2017experiments,ilderton2016prospects}.

Recent progress of high-power laser and electron accelerator technologies have stimulated the interest of making laser-driven ultrarelativistic polarized particle sources via nonlinear Compton scattering \cite{li2019ultrarelativistic,li2020polarized,chen2019polarized,song2019spin,
wan2020ultrarelativistic,dai2022photon,li2020production}. 
By shining laser light at a counter-propagating ultrarelativistic electrons beam, the electron spin varies due to spin precession \cite{thomas1926motion,thomas1927kinematics,bargmann1959precession,walser2002spin}, radiative polarization \cite{Baier1972,Seipt2018,Karlovets2011} and QED loop effects \cite{Baier1972,Meuren2011,torgrimsson2021resummation,Torgrimsson_2021,ilderton2020loop}. After the interaction, the electrons could obtain a net polarization along a certain direction in the case that the background fields have some sort of asymmetry such as the elliptically polarized \cite{li2019ultrarelativistic,li2020polarized} and two-color laser fields \cite{chen2019polarized}. The asymmetry is essential to prevent the cancellation of polarization effects in the adjacent half-cycles.

To investigate the electron spin dynamics and photon polarization in strong laser fields, the spin-resolved simulation approaches have been developed using probabilistic routine based on electron spin- and photon polarization-resolved quantum radiation probability \cite{li2019ultrarelativistic,li2020polarized,chen2022electron}. Under local constant field approximation (LCFA)    \cite{di2019improved,seipt2020spin,ilderton2019extended,lv2021anomalous,ritus1985quantum,
king2020nonlinear,di2012extremely,seipt2018theory}, the quantum probability can be obtained using strong-field QED in the Furry picture \cite{Torgrimsson_2021,dinu2020approximating,seipt2020spin,king2020nonlinear,
wistisen2014interference,mackenroth2011nonlinear} or quantum operator method introduced by Baier and Katkov \cite{chen2022electron,katkov1998electromagnetic}, which are equivalent under constant cross field approximation. However, the spin-resolved probability applied in the strong-field simulation codes are based on integration over emitted photon angle \cite{chen2022electron}. Consequently, the angular spread around the electron propagation direction at the emission events have to be ignored, as well as the dependency of polarization on the emission angle. This simplification is justified in the ultrarelativistic limit as the angular spread at emission (ASE) $\Delta\theta\sim 1/\gamma\ll1$, much smaller than the total angle of particle deflection in the intense laser field $\theta_D=a_0/\gamma$. 
However, whether the approximation is valid in a realistic polarization scenario haven't been truly tested. 
The effects of ASE on dynamics and polarization of the out-going particles could be important, especially for the low-energy photon emissions in the laser with $a_{0i}=\bm{a}_{0}\cdot \mathbf{\hat{e}}_i\ll 1$ along a certain direction $\mathbf{\hat{e}}_i$. Here $\bm{a}_0=|e|\bm{E}_0/m\omega_0$ is the invariant laser field parameter, with $-e$ and $m$ being the electron charge and mass, respectively, $E_0$ and $\omega_0$ the amplitude and frequency of the laser field, respectively. In this case, the deflection angle is negligible along $\mathbf{\hat{e}}_i$, while the ASE is nontrivial due to the smallness of $\gamma$.

The exact angle-resolved radiation probability can be calculated with semi-classical or volkov-state approaches, which is valid for arbitrary electron spin and photon polarization \cite{berestetskii1982quantum,wistisen2019numerical,wistisen2020numerical,thomas2010algorithm}. However, the radiation probability obtained in these literatures include double integrals over interaction phase and are numerically cumbersome in the strong field regime, where formation length is small and multiple emissions dominate. Under local constant field approximation (LCFA), the angle-resolved probability has been recently derived after averaging over initial spins and summing over final spins of electrons, and applied to study the energy and angular spectra of polarized photon in the weakly nonlinear regime \cite{king2020nonlinear}. The angle-resolved LCFA probability for unpolarized photons can be found in \cite{katkov1998electromagnetic}, which has been applied for investigating radiation beaming in the quantum regime without polarizations \cite{blackburn2020radiation}.

In this paper, we derived the angle-resolved probability for nonlinear Compton scattering using the Baier-Katkov quantum operator method under 
LCFA, valid for arbitrary electrons spin and photon polarization.
By applying the probability to the fully spin-resolved QED codes, we developed an angle-resolved Monte-Carlo method for investigating the effect of ASE on angular distribution of density and polarization for out-going particles in arbitrary laser fields. We revisited the polarization schemes in an elliptically polarized laser field with the newly developed simulation method. 
We show that the ASE could cause a remarkable spread of angular distribution along the direction with negligible deflection angle ($\theta_D^i\propto\bm{a}_0\cdot\mathbf{\hat{e}}_i\ll1$), and affected the direction of radiation reaction force, which consequently decrease the angle-dependent polarization of final electrons.

\section{Angle-resolved LCFA probability }

\begin{figure}
    \includegraphics[width=0.45\textwidth]{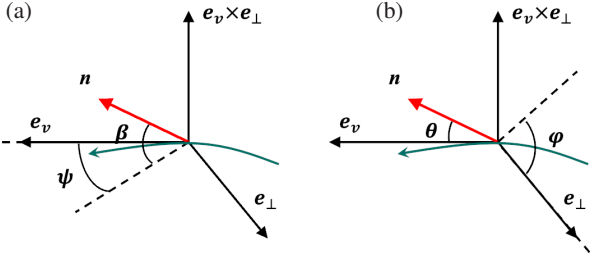}
     \begin{picture}(300,20)
    \put(10,110){(a)}
    \put(150,110){(b)}
	\end{picture}
    \caption{Reference frames for description of the angular characteristics of the radiation. (a) $\beta$ is an angle between the plane $(\mathbf{e}_v,\mathbf{e}_\perp)$ and vector $\bf{n}$, and $\psi$ is an angle between the projection of vector $\bf{n}$ on the plane $(\mathbf{e}_v,\mathbf{e}_\perp)$ and vector $\mathbf{e}_v$. (b) $\theta$ is the angle between $\bf{n}$  and $\mathbf{e}_v$, and $\varphi$ is the angle between the projection of vector $\bf{n}$ and vector $\mathbf{e}_\perp$. }
    \label{Fig.agl}
\end{figure}
The electrons in the laser fields radiate mainly forward into the narrow cone with the axis directed along the electron velocity $\mathbf{v}$ and the cone opening angle $\Delta \theta\sim1/\gamma$. The radiation direction $\mathbf{n}=(\cos\beta\cos\psi,\cos\beta\sin\psi,\sin\beta)$ is expressed with the emission angles $(\beta,\psi)$ [Fig. \ref{Fig.agl} (a)], in the basis spanned by 3 orthogonal vectors $(\mathbf{e}_v,\mathbf{e}_\perp,\mathbf{e}_v\times\mathbf{e}_\perp)$ with $\mathbf{e}_v=\mathbf{v}/|\mathbf{v}|$ being the unit vector along the electron velocity $\mathbf{v}(t)$, $\mathbf{e}_\perp=\bf{w}_\perp/|\bf{w}_\perp|$ the unit vector along the direction of the transverse component of acceleration $\mathbf{w}_\perp(t)$ [Fig. \ref{Fig.agl}]. 
The emission probability for a photon radiated along $\mathbf{n}$ can obtained by Baier-Katkov quantum operator
method \cite{katkov1998electromagnetic}

\begin{equation}\label{PRB}
dw_{rad}=\frac{\alpha}{\left(2\pi\right)^{2}}\frac{d^{3}\mathbf{k}}{\omega}\int dt_{1}\int dt_{2}R_{2}^{*}R_{1}\exp\left[-i\frac{\varepsilon\left(kx_{2}-kx_{1}\right)}{\varepsilon'}\right],
\end{equation}
where $k^{\mu}=\omega\left\{ 1,\mathbf{n}\right\} $ and $x^{\mu}=\left\{ t,\mathbf{r}(t)\right\} $
are the 4-momentum and 4-coordinate of the emitted photon. The indices
1 and 2 denote the dependence on the radiation time moments $t_{1}$
and $t_{2}$ along the radiation direction $\mathbf{n}$, respectively. $\varepsilon$ and $\varepsilon'$ are the electron
energies before and after emission, respectively, and
\begin{align}
\label{R}
R(t) & =\varphi'^{+}(\bm{\zeta}')\left[A(t)+i\bm{\sigma}\cdot \mathbf{B}(t)\right]\varphi(\bm{\zeta}),
\end{align}
where $\varphi$ and $\varphi'$ are the two-component spinors
that describe the initial and final spin states of the electron,
respectively. The unit vectors $\bm{\zeta}$ and $\bm{\zeta}'$
are the corresponding spin vectors.
The expressions of $A(t)$ and $\mathbf{B}(t)$ are
\begin{align}\label{AB}\nonumber
A(t) & =\frac{\mathbf{e}^{*}\cdot\mathbf{p}(t)}{2\sqrt{\varepsilon\varepsilon'}}\left[\left(\frac{\varepsilon'+m}{\varepsilon+m}\right)^{1/2}+\left(\frac{\varepsilon+m}{\varepsilon'+m}\right)^{1/2}\right],\\ \nonumber
\mathbf{B}(t)&=\frac{1}{2\sqrt{\varepsilon\varepsilon'}}\left[\left(\frac{\varepsilon'+m}{\varepsilon+m}\right)^{1/2}\mathbf{e}^{*}\times\mathbf{p}(t)+\left(\frac{\varepsilon+m}{\varepsilon'+m}\right)^{1/2}\mathbf{e}^{*}\times\right.\\
&\left.\left(\mathbf{p}(t)-\mathbf{k}\right)\right],
\end{align}
with $\mathbf{p}(t)=\gamma m \mathbf{v}(t) $ being the momentum of the
electron, $\gamma=\varepsilon/m$ the Lorenz factor, $\mathbf{e}$
the polarization vector of the emitted photon, which can be expressed in terms of the orthogonal basis
\begin{eqnarray}
\mathbf{e}_1&=&\frac{\mathbf{s}-(\mathbf{n}\cdot\mathbf{s})\mathbf{n}}{|\mathbf{s}-(\mathbf{n}\cdot\mathbf{s})\mathbf{n}|}, \quad \mathbf{e}_2=\left[\mathbf{n}\times \mathbf{s}\right],
\end{eqnarray}
where $\bf{s}=\bf{w}/|\bf{w}|$. Changing variables from $t_1$, $t_2$ to $t=\left(t_{1}+t_{2}\right)/2$ and $\tau=t_{2}-t_{1}$, the LCFA result can be obtained by expanding $\mathbf{v}_{1,2}$ and $\mathbf{r}_{1,2}$ in Eq. (\ref{PRB})  over $\tau$ \cite{katkov1998electromagnetic,chen2022electron}. Keeping in mind that  $\beta$ and $\psi$ are both in the order of $\sim 1/\gamma$, the combinations of angles with order beyond $\sim 1/\gamma^2$ can be dropped out. Then one could obtain the photon radiation probability per unit time and per solid angle with the accuracy up to $\sim1/\gamma^2$:
\begin{eqnarray}\label{PRB2}
\frac{d^2W_{rad}}{d\omega d\Omega}&=&\frac{\alpha\omega}{\left(2\pi\right)^{2}} \int_{-\infty}^\infty d\tau R_{2}^{*}R_{1}\\ \nonumber
&&\exp\left\{ -i\frac{\varepsilon}{\varepsilon'}\omega\left[\left(\frac{\beta^{2}}{2}+\frac{\psi^{2}}{2}+\frac{1}{2\gamma^{2}}\right)\tau+\frac{w^{2}\tau^{3}}{24}\right]\right\}.
\end{eqnarray}
After the integration over $\tau$, 
we obtain the polarization matrix of radiation probability per unit time and per solid angle:

\begin{widetext}
\begin{align*}
\frac{d^2W_{11}+d^2W_{22}}{d\omega d\Omega}&=C\left\{ \left[\frac{\varepsilon^{2}-\varepsilon'^{2}}{\varepsilon'\varepsilon}\theta\sin\varphi\left(\boldsymbol{\zeta}+\boldsymbol{\zeta}'\right)\cdot\hat{\mathbf{v}}-\frac{1}{\gamma}\left(\frac{\omega}{\varepsilon}\boldsymbol{\zeta}+\frac{\omega}{\varepsilon'}\boldsymbol{\zeta}'\right)\cdot\mathbf{b}\right]\Theta^{\frac{1}{2}}\textrm{K}_{\frac{2}{3}}\left(\xi\right)\right.\\&+\left\{ \theta^{2}+\Theta\left(\frac{\varepsilon^{2}+\varepsilon'^{2}}{\varepsilon'\varepsilon}-1\right)+\left[\frac{\varepsilon^{2}+\varepsilon'^{2}}{2\varepsilon'\varepsilon}\theta^{2}+\frac{\omega^{2}}{2\varepsilon'\varepsilon}\frac{1}{\gamma^{2}}+\left(1-\frac{\omega^{2}}{2\varepsilon'\varepsilon}\right)\Theta\right]\ensuremath{\left(\boldsymbol{\zeta}\cdot\boldsymbol{\zeta}'\right)}\right\} \textrm{K}_{\frac{1}{3}}\left(\xi\right)\\&+\frac{\varepsilon^{2}-\varepsilon'^{2}}{2\varepsilon'\varepsilon}\left[\cos\varphi\mathbf{b}-\sin\varphi\mathbf{s}\right]\cdot\left[\boldsymbol{\zeta}'\times\boldsymbol{\zeta}\right]\frac{1}{\gamma}\theta\textrm{K}_{\frac{1}{3}}\left(\xi\right)\\\mathbf{}&+\frac{\omega^{2}}{2\varepsilon'\varepsilon}\theta\textrm{K}_{\frac{1}{3}}\left(\xi\right)\left\{ 2\theta\left(\boldsymbol{\zeta}\cdot\hat{\mathbf{v}}\right)\left(\boldsymbol{\zeta}'\cdot\hat{\mathbf{v}}\right)-\frac{\cos\varphi}{\gamma}\left[\left(\boldsymbol{\zeta}\cdot\mathbf{s}\right)\left(\boldsymbol{\zeta}'\cdot\hat{\mathbf{v}}\right)+\left(\boldsymbol{\zeta}\cdot\hat{\mathbf{v}}\right)\left(\boldsymbol{\zeta}'\cdot\mathbf{s}\right)\right]\right.\\&\left.\left.-\frac{\sin\varphi}{\gamma}\left[\left(\boldsymbol{\zeta}\cdot\mathbf{b}\right)\left(\boldsymbol{\zeta}'\cdot\hat{\mathbf{v}}\right)+\left(\boldsymbol{\zeta}\cdot\hat{\mathbf{v}}\right)\left(\boldsymbol{\zeta}'\cdot\mathbf{b}\right)\right]\right\} \right\},
\end{align*}

\begin{align*}
\frac{d^2W_{12}+d^2W_{21}}{d\omega d\Omega}&=C\left\{ \left[\frac{\omega}{\varepsilon'}\left(\boldsymbol{\zeta}\cdot\mathbf{s}\right)+\frac{\omega}{\varepsilon}\left(\boldsymbol{\zeta}'\cdot\mathbf{s}\right)\right]\frac{1}{\gamma}\Theta{}^{\frac{1}{2}}\textrm{K}_{\frac{2}{3}}\left(\xi\right)\right.\\&+\theta^{2}\sin2\varphi\left[1+\frac{\varepsilon^{2}+\varepsilon'^{2}}{2\varepsilon'\varepsilon}\textrm{\ensuremath{\left(\boldsymbol{\zeta}\cdot\boldsymbol{\zeta}'\right)}}-\frac{\omega^{2}}{2\varepsilon'\varepsilon}\left(\boldsymbol{\zeta}\cdot\hat{\mathbf{v}}\right)\left(\boldsymbol{\zeta}'\cdot\hat{\mathbf{v}}\right)\right]\textrm{K}_{\frac{1}{3}}\left(\xi\right)\\&+\frac{\varepsilon^{2}-\varepsilon'^{2}}{2\varepsilon'\varepsilon}\left\{ \left(\theta^{2}\cos2\varphi+\Theta\right)\mathbf{\hat{\mathbf{v}}}-\frac{\theta\cos\varphi}{\gamma}\mathbf{s}+\frac{\theta\sin\varphi}{\gamma}\mathbf{b}\right\} \cdot\left[\boldsymbol{\zeta}'\times\boldsymbol{\zeta}\right]\textrm{K}_{\frac{1}{3}}\left(\xi\right)\\&+\frac{\omega^{2}}{2\varepsilon'\varepsilon}\textrm{K}_{\frac{1}{3}}\left(\xi\right)\left\{ -\frac{1}{\gamma^{2}}\left[\left(\boldsymbol{\zeta}\cdot\mathbf{b}\right)\left(\boldsymbol{\zeta}'\cdot\mathbf{s}\right)+\left(\boldsymbol{\zeta}\cdot\mathbf{s}\right)\left(\boldsymbol{\zeta}'\cdot\mathbf{b}\right)\right]\right.\\&\left.\left.+\frac{\theta\sin\varphi}{\gamma}\left[\left(\boldsymbol{\zeta}\cdot\hat{\mathbf{v}}\right)\left(\boldsymbol{\zeta}'\cdot\mathbf{s}\right)+\left(\boldsymbol{\zeta}\cdot\mathbf{s}\right)\left(\boldsymbol{\zeta}'\cdot\hat{\mathbf{v}}\right)\right]+\frac{\theta\cos\varphi}{\gamma}\left[\left(\boldsymbol{\zeta}\cdot\hat{\mathbf{v}}\right)\left(\boldsymbol{\zeta}'\cdot\mathbf{b}\right)+\left(\boldsymbol{\zeta}\cdot\mathbf{b}\right)\left(\boldsymbol{\zeta}'\cdot\hat{\mathbf{v}}\right)\right]\right\} \right\},
\end{align*}

\begin{align*}
\frac{d^2W_{12}-d^2W_{21}}{d\omega d\Omega}&=iC\left\{ \left\{ \frac{\varepsilon^{2}-\varepsilon'^{2}}{2\varepsilon'\varepsilon}\frac{1}{\gamma}\left(\mathbf{s}\cdot\left[\boldsymbol{\zeta}'\times\boldsymbol{\zeta}\right]\right)-\theta\sin\varphi\left[\frac{\varepsilon^{2}+\varepsilon'^{2}}{\varepsilon'\varepsilon}+2\left(\boldsymbol{\zeta}\cdot\boldsymbol{\zeta}'\right)\right]\right\} \Theta^{\frac{1}{2}}\textrm{K}_{\frac{2}{3}}\left(\xi\right)\right.\\&+\left[\left(\frac{\omega}{\varepsilon}\frac{1}{\gamma^{2}}-\frac{\varepsilon^{2}-\varepsilon'^{2}}{\varepsilon'\varepsilon}\Theta\right)\hat{\mathbf{v}}+\frac{\omega}{\varepsilon}\frac{\theta}{\gamma}\left(\cos\varphi\mathbf{s}+\sin\varphi\mathbf{b}\right)\right]\cdot\boldsymbol{\zeta}\textrm{K}_{\frac{1}{3}}\left(\xi\right)\\&+\left[\left(\frac{\omega}{\varepsilon^{'}}\frac{1}{\gamma^{2}}-\frac{\varepsilon^{2}-\varepsilon'^{2}}{\varepsilon'\varepsilon}\Theta\right)\hat{\mathbf{v}}+\frac{\omega}{\varepsilon'}\frac{\theta}{\gamma}\left(\cos\varphi\mathbf{s}+\sin\varphi\mathbf{b}\right)\right]\cdot\boldsymbol{\zeta}'\textrm{K}_{\frac{1}{3}}\left(\xi\right)\\&\left.-\frac{\omega^{2}}{\varepsilon'\varepsilon}\Theta^{\frac{1}{2}}\textrm{K}_{\frac{2}{3}}\left(\xi\right)\left\{ \theta\sin\varphi\left(\boldsymbol{\zeta}\cdot\hat{\mathbf{v}}\right)\left(\boldsymbol{\zeta}'\cdot\hat{\mathbf{v}}\right)-\frac{1}{2\gamma}\left[\left(\boldsymbol{\zeta}\cdot\hat{\mathbf{v}}\right)\left(\boldsymbol{\zeta}'\cdot\mathbf{b}\right)+\left(\boldsymbol{\zeta}\cdot\mathbf{b}\right)\left(\boldsymbol{\zeta}'\cdot\hat{\mathbf{v}}\right)\right]\right\} \right\},
\end{align*}

\begin{align}\nonumber\label{ARP}
\frac{d^2W_{11}-d^2W_{22}}{d\omega d\Omega}&=C\left\{ -\left[\frac{\omega}{\varepsilon}\left(\boldsymbol{\zeta}'\cdot\mathbf{b}\right)+\frac{\omega}{\varepsilon^{'}}\left(\boldsymbol{\zeta}\cdot\mathbf{b}\right)\right]\frac{1}{\gamma}\Theta{}^{\frac{1}{2}}\textrm{K}_{\frac{2}{3}}\left(\xi\right)\right.\\\nonumber
&+\left[\theta^{2}\cos2\phi+\Theta\right]\left[1+\left(\boldsymbol{\zeta}\cdot\boldsymbol{\zeta}'\right)\frac{\varepsilon^{2}+\varepsilon'^{2}}{2\varepsilon'\varepsilon}\right]\textrm{K}_{\frac{1}{3}}\left(\xi\right)\\\nonumber
&+\frac{\varepsilon^{2}-\varepsilon^{'2}}{2\varepsilon^{'}\varepsilon}\left[-\theta^{2}\sin2\varphi\mathbf{\hat{\mathbf{v}}}+\frac{\theta\sin\varphi}{\gamma}\mathbf{s}+\frac{\theta\cos\varphi}{\gamma}\mathbf{b}\right]\cdot\left[\boldsymbol{\zeta}'\times\boldsymbol{\zeta}\right]\textrm{K}_{\frac{1}{3}}\left(\xi\right)\\\nonumber
&+\frac{\omega^{2}}{2\varepsilon^{'}\varepsilon}\textrm{K}_{\frac{1}{3}}\left(\xi\right)\left\{ -\left(\theta^{2}\cos2\varphi+\Theta\right)\left(\boldsymbol{\zeta}\cdot\hat{\mathbf{v}}\right)\left(\boldsymbol{\zeta}'\cdot\hat{\mathbf{v}}\right)+\frac{1}{\gamma^{2}}\left[\left(\boldsymbol{\zeta}\cdot\mathbf{b}\right)\left(\boldsymbol{\zeta}'\cdot\mathbf{b}\right)-\left(\boldsymbol{\zeta}\cdot\mathbf{s}\right)\left(\boldsymbol{\zeta}'\cdot\mathbf{s}\right)\right]\right.\\&\left.\left.+\frac{\theta\cos\varphi}{\gamma}\left[\left(\boldsymbol{\zeta}\cdot\mathbf{s}\right)\left(\boldsymbol{\zeta}'\cdot\hat{\mathbf{v}}\right)+\left(\boldsymbol{\zeta}\cdot\hat{\mathbf{v}}\right)\left(\boldsymbol{\zeta}'\cdot\mathbf{s}\right)\right]-\frac{\theta\sin\varphi}{\gamma}\left[\left(\boldsymbol{\zeta}\cdot\hat{\mathbf{v}}\right)\left(\boldsymbol{\zeta}'\cdot\mathbf{b}\right)+\left(\boldsymbol{\zeta}\cdot\mathbf{b}\right)\left(\boldsymbol{\zeta}'\cdot\hat{\mathbf{v}}\right)\right]\right\} \right\}.
\end{align}
\end{widetext}
Where $\mathbf{\hat{v}}=\mathbf{v}/\left|\mathbf{v}\right|$,
$\mathbf{b}=\mathbf{\hat{v}}\times\mathbf{s}$, $C=\frac{\alpha\gamma^{3}\omega\Theta^{\frac{1}{2}}}{2\sqrt{3}\pi^{2}\chi\varepsilon'}$ with $\Theta=\theta^{2}+\frac{1}{\gamma^{2}}$, $\xi=\frac{2}{3}\frac{\omega}{\varepsilon'\chi}\left[2\gamma^{2}\left(1-nv\right)\right]^{\frac{3}{2}}=\frac{2}{3}\frac{\omega}{\varepsilon'\chi}\left(\gamma^{2}\Theta\right)^{\frac{3}{2}}$. Note that, we have convert the results to the  $(\theta,\varphi)$ frame [Fig. \ref{Fig.agl} (b)], for the sake of convenience of further integration of emission angles and obtaining the angle-unresolved LCFA probability in \cite{chen2022electron}.
The angle-resolved radiation probability density including all the polarization and spin characteristic takes the form

\begin{align}\label{PRB_tot_angle}
 \frac{d^2W_{rad}}{d\omega d\Omega} &=\frac{1}{2}\left(F_0+\xi_1F_1+\xi_2F_2+\xi_3F_3\right),
\end{align}
where $F_0= \frac{d^2W_{11}+ d^2W_{22}}{d\omega d\Omega}$, $F_1\frac{d^2W_{12}+ d^2W_{21}}{d\omega d\Omega}$, $F_2=i\frac{d^2W_{12}- d^2W_{21}}{d\omega d\Omega}$, $F_3=\frac{d^2W_{11}- d^2W_{22}}{d\omega d\Omega}$,
and the 3-vector $\bm{\xi}=\left(\xi_1,\xi_2,\xi_3\right)$ is the Stokes parameters of emitted photon defined with respect to $\mathbf{e}_1 $ and $\mathbf{e}_2$.
For an arbitrarily polarized photon with polarization vector $\mathbf{e}=a_{1}\mathbf{e}_1+a_{2}\mathbf{e}_2$, Stokes parameters are given by
\begin{align}
\xi_{1}=a_{1}a_{2}^{\ast}+a_{2}a_{1}^{\ast};\;\xi_{2}=i\left(a_{1}a_{2}^{\ast}-a_{2}a_{1}^{\ast}\right);\;\xi_{3}=\left|a_{1}\right|^{2}-\left|a_{2}\right|^{2}
\end{align}
\begin{figure}[]
    \includegraphics[width=0.45\textwidth]{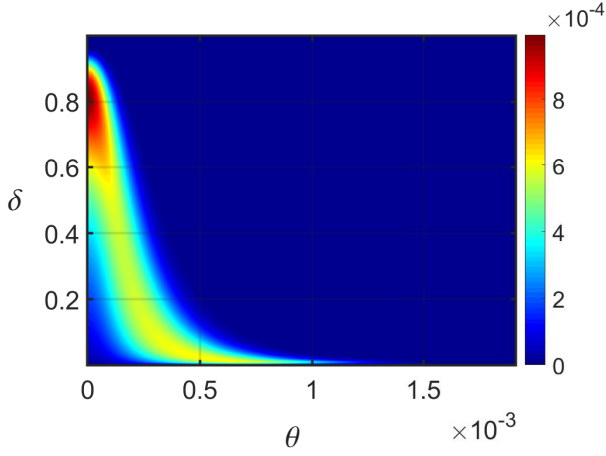}
    \begin{picture}(300,20)
    \put(110,20){\large$\theta$}
	\put(5,110){\large$\delta$}
    \end{picture}
    \caption{The angle-resolved radiation probability $\frac{d^2\widetilde{W}_{rad}}{d\omega d\Omega}$ vs emitted photon energy $\delta=\omega/\varepsilon$ and emission angle $\theta$ for $\chi=2.0$. }
    \label{Fig.delta_theta}
\end{figure}

\begin{figure*}[t]
    \includegraphics[width=0.9\textwidth]{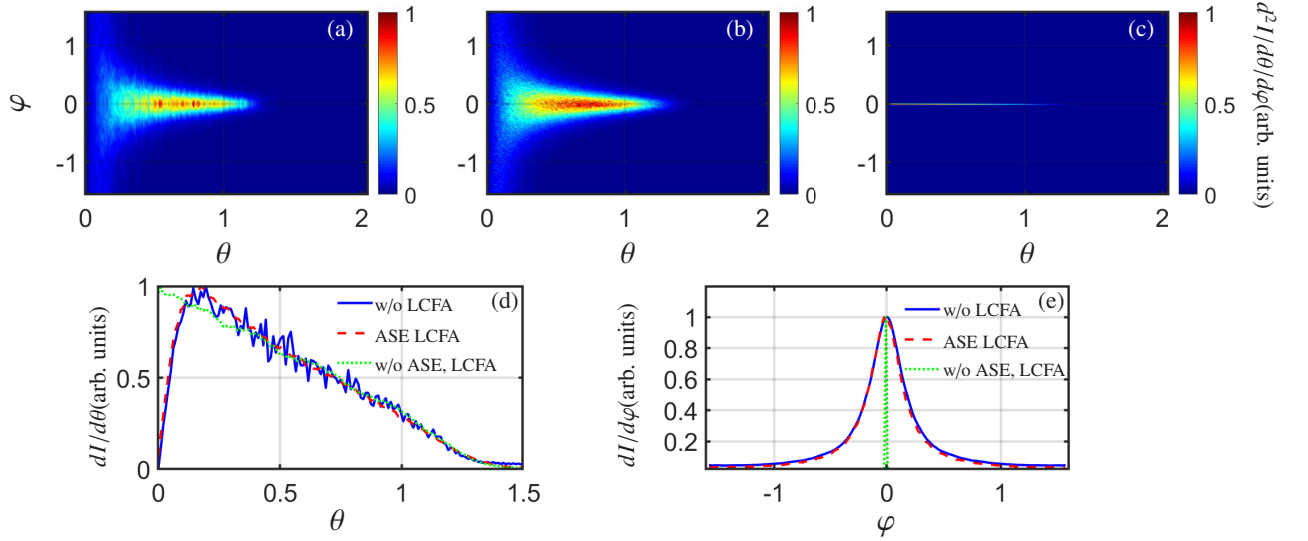}
    \begin{picture}(300,20)
    \put(380,230){\rotatebox{-90}{ $d^2I/d\theta/d\varphi$(arb. units)}}
	\put(30,215){\color{white}(a)}
	\put(181,215){\color{white}(b)}
	\put(333,215){\color{white}(c)}
	\put(-12,129){\large$\theta$}
	\put(141,129){\large$\theta$}
	\put(292,129){\large$\theta$}
	\put(-90,186){\rotatebox{90}{\large$\varphi$}}
    \put(92,112){(d)}
    \put(-60,52){\rotatebox{90}{$dI/d\theta$(arb. units)}}
    \put(31,28){\large$\theta$}
    \put(48,111){\scriptsize w/o LCFA}
    \put(48,99){\scriptsize ASE LCFA }
    \put(48,88){\scriptsize w/o ASE, LCFA}
    \put(299,112){(e)}
    \put(238,28){\large$\varphi$}
    \put(140,52){\rotatebox{90}{$dI/d\varphi$(arb. units)}}
    \put(263,109){\scriptsize w/o LCFA}
    \put(263,97){\scriptsize ASE LCFA  }
    \put(263,86){\scriptsize w/o ASE, LCFA }
	\end{picture}
    \caption{The angular intensity of emitted photons $d^2I/d\theta/d\varphi$ versus $\theta$ (mrad) and $\varphi$ (mrad): (a) for semiclassical approach without LCFA, (b) for angle-resolved LCFA approach, and (c) for angle-unresolved LCFA approach.  The angular intensity of photons $dI/d\theta$ vs $\theta$  (mrad) (d)  and  $dI/d\varphi $ vs $\varphi$  (mrad) (e) for semiclassical approach without LCFA (blue solid line), angle-resolved LCFA approach (red dashed line) and angle-unresolved LCFA approach (green dotted line)}.
    \label{Fig.compare}
\end{figure*}
After summing over the  polarization of emitted photon, we get

\begin{widetext}
\begin{align}\nonumber
\frac{d^2\overline{W}_{rad}\left(\boldsymbol{\zeta},\boldsymbol{\zeta}'\right)}{d\omega d\Omega}&=C\left(\alpha+\boldsymbol{\eta}\cdot\boldsymbol{\zeta}'\right)\\\nonumber\alpha&=\left[\theta^{2}+\Theta\left(\frac{\varepsilon^{2}+\varepsilon'^{2}}{\varepsilon'\varepsilon}-1\right)\right]\textrm{K}_{\frac{1}{3}}\left(\xi\right)+\Theta{}^{\frac{1}{2}}\left[\frac{\varepsilon^{2}-\varepsilon'^{2}}{\varepsilon'\varepsilon}\theta\sin\varphi\left(\boldsymbol{\zeta}\cdot\hat{\mathbf{v}}\right)-\frac{\omega}{\varepsilon}\frac{1}{\gamma}\left(\boldsymbol{\zeta}\cdot\mathbf{b}\right)\right]\textrm{K}_{\frac{2}{3}}\left(\xi\right)\\\nonumber\boldsymbol{\eta}&=\left(\frac{\varepsilon^{2}-\varepsilon'^{2}}{\varepsilon'\varepsilon}\theta\sin\varphi\hat{\mathbf{v}}-\frac{\omega}{\varepsilon'}\frac{1}{\gamma}\mathbf{b}\right)\Theta{}^{\frac{1}{2}}\textrm{K}_{\frac{2}{3}}\left(\xi\right)\\\nonumber&+\left\{ \left[\frac{\varepsilon^{2}+\varepsilon'^{2}}{2\varepsilon'\varepsilon}\theta^{2}+\frac{\omega^{2}}{2\varepsilon'\varepsilon}\frac{1}{\gamma^{2}}+\left(1-\frac{\omega^{2}}{2\varepsilon'\varepsilon}\right)\Theta\right]\boldsymbol{\zeta}+\frac{\varepsilon^{2}-\varepsilon'^{2}}{2\varepsilon'\varepsilon}\frac{1}{\gamma}\theta\boldsymbol{\zeta}\times\left(\cos\varphi\mathbf{b}-\sin\varphi\mathbf{\mathbf{s}}\right)\right\} \textrm{K}_{\frac{1}{3}}\left(\xi\right)\\&+\left\{ 2\theta^{2}\left(\boldsymbol{\zeta}\cdot\hat{\mathbf{v}}\right)\hat{\mathbf{v}}-\frac{\theta\cos\varphi}{\gamma}\left[\left(\boldsymbol{\zeta}\cdot\mathbf{s}\right)\hat{\mathbf{v}}+\left(\boldsymbol{\zeta}\cdot\hat{\mathbf{v}}\right)\mathbf{s}\right]-\frac{\theta\sin\varphi}{\gamma}\left[\left(\boldsymbol{\zeta}\cdot\mathbf{b}\right)\hat{\mathbf{v}}+\left(\boldsymbol{\zeta}\cdot\hat{\mathbf{v}}\right)\mathbf{b}\right]\right\} \frac{\omega^{2}}{2\varepsilon'\varepsilon}\textrm{K}_{\frac{1}{3}}\left(\xi\right).
\end{align}
\end{widetext}
The final polarization vector of the electron resulting from the scattering process itself is $\bm{\zeta}^f = \frac{\boldsymbol{\eta}}{\alpha}$ \cite{jackson1999classical}. After averaging over initial and summing over final electron polarizations, we get the analytic expression for the angle-resolved probability density:
\begin{align}\label{delta_theta}
\frac{d^2\widetilde{W}_{rad}}{d\omega d\Omega}&=2C\left[\theta^{2}+\Theta\left(\frac{\varepsilon^{2}+\varepsilon'^{2}}{\varepsilon'\varepsilon}-1\right)\right]\textrm{K}_{\frac{1}{3}}\left(\xi\right).
\end{align}
The angular spread of a photon around the emitting electron is inversely proportional to emitted photon energy $\delta$ [see Fig. \ref{Fig.delta_theta}], roughly with $\theta\sim \delta^{-1/3}$ for $\delta\ll1$ \cite{jackson1999classical} and $\theta\sim \sqrt{(1-\delta)/\delta}$ for $\delta\sim1$ \cite{blackburn2020radiation}. Therefore, the ASE effects is more significant for low-energy photon emissions. Let us compare our results with the angle-resolved LCFA results obtained earlier. 
The angle-resolved probability given in Eq. (\ref{delta_theta}) is same with Eq. (4.23) in \cite{katkov1998electromagnetic} and Eq. (1) in \cite{blackburn2020radiation}, using the relation $u=\omega/\varepsilon'$ and $z=\left[\gamma^{2}\theta^{2}+1\right]^{3/2}$.

\section{simulation result}
\subsection{Angle-resolved Monte-Carlo simulation}
Let us first introduce our angle-resolved Monte Carlo method for nonlinear Compton scattering. Photon emissions are treated quantum mechanically, while the electron dynamics semiclassically. At each simulation step, two random numbers are sampled to decide the occurrence of photon emission and photon energy by the spectral probability. The spin of the electron after the emission is determined by the spin-resolved emission probabilities according to the commonly used stochastic algorithm \cite{li2020polarized,chen2019polarized,song2019spin,wan2020high}, as well as the polarization of the emitted photons \cite{li2020polarized,dai2022photon}. Once photon energy $\omega$, electron spin $\bm{\zeta}_f$ and Stokes parameters $\bm{\xi}$ have been picked, the angles of emission are picked using the distribution $d^2W_{rad}/d\omega d\Omega$ of Eq. (\ref{PRB_tot_angle}) and the commonly used acceptance-rejection method. 
After emission, the electron momentum is changed according to the momentum conservation, $\bm{p}'=\bm{p}-\bm{k}(\theta,\varphi)$, enabling an angle-resolved radiation reaction.  In this way, an angular corrections to the kinetic dynamics of all the put-going particles is included to the strong-field QED simulation codes.

To confirm the accuracy of our simulation method, we simulated the nonlinear Compton scattering of a linearly polarized laser pulse and ultrarelativistic electron beam with different approaches, including the semiclassical approach without LCFA  \cite{baier1968quasiclassical,katkov1998electromagnetic,berestetskii1982quantum}, the angle-resolved and -unresolved LCFA approaches [see Fig. \ref{Fig.compare}]. The analytical expression for the spectral distribution of radiation can be derived in the framework of the Baier–Katkov semiclassical approximation based on the classical trajectory \cite{baier1968quasiclassical,katkov1998electromagnetic,berestetskii1982quantum,belkacem1985theory}.  With the semiclassical approach, one could obtain the angular distribution of emitted photons by coherently integrating over the electron trajectory, which has been proved in \cite{wistisen2014interference,wistisen2019numerical} to be coincide with the exact QED calculations. 
However, this method is more suitable for investigating the interference effects or polarization effects with moderate laser intensity, where formation length is comparable to the timescale of the external field and multiple photon emissions is negligible. In the case that the formation length is much smaller than the field inhomogeneities, LCFA can be adopted to avoid the complicated integrals in simulation. Therefore, we compared the semiclassical approach with integration of classical trajectory and angle-resolved LCFA approach in the single-photon emission dominated regime, to assess the validity of LCFA and capacity of describing angular distribution. Meanwhile, we also compared the angle-resolved approaches and angle-unresolved LCFA approach to reveal the importance of the angle spread at emissions.

The angular distribution of emitted photon energy with different approaches are shown in Fig. \ref{Fig.compare}. We considered a linearly polarized (along $x$ axis) planewave, with peak intensity $I\approx 10^{21}\text{W/cm}^2 (a_0=10)$, wavelength of the laser  $\lambda_0=1.0\mu m$, pulse duration $\tau_p=5T_0$ with period $T_0$. The parameters are chosen such that multiple emissions can be avoided. The average photon number emitted by a single electrons can be estimated by $N_{\gamma}\sim\alpha a_{0}\tau_{p}/T_{0}< 1$. 
For the semiclassical approach without LCFA,  we obtain the trajectory of an 4-GeV electron by solving Lorentz equation, and then substitute the time-dependent momentum and coordinate of the electron into the radiation spectrum \cite{belkacem1985theory,wistisen2014interference}:
\begin{equation}
\frac{d^{2}I}{d\Omega d\omega}=\frac{e^{2}}{4\pi^{2}}\left(\frac{\varepsilon^{2}+\varepsilon'^{2}}{2\varepsilon^{2}}\left|\bm{I}\right|^{2}+\frac{\omega^{2}m^{2}}{2\varepsilon^{4}}\left|J\right|^{2}\right),
\end{equation}
where
\begin{align*}
\bm{I}&=\int\frac{\mathbf{n}\times\left[\left(\mathbf{n}-\mathbf{v}\right)\times\dot{\mathbf{v}}\right]}{\left(1-\mathbf{n}\cdot\mathbf{v}\right)^{2}}e^{i\frac{\varepsilon}{\varepsilon'}kx}dt,\\
J&=\int\frac{\mathbf{n}\cdot\dot{\mathbf{v}}}{\left(1-\mathbf{n}\cdot\mathbf{v}\right)^{2}}e^{i\frac{\varepsilon}{\varepsilon'}kx}dt.
\end{align*}
As shown in Fig. \ref{Fig.compare} (a), the azimuthal angle of photons emitted by an electron in the linearly polarized laser are centred at $\varphi=\text{tan}^{-1}p_y/p_x=0$ with a width of $\Delta \varphi=0.4$rad, and polar angle $\theta=\text{cos}^{-1}p_z/|p|$ is within $\sim a_0/\gamma=0.5$mrad. 

For the LCFA approaches, we consider a monochromatic electron beam consists of $3\times10^6$ electrons, which has a cylindrical form with radius $w_e=\lambda_0$, length $L_e=5\lambda_0$ and initial energy of $\varepsilon_0=4$GeV. The electron density has a transversely Gaussian and longitudinally uniform distribution. The simulation results show that the angle-resolved LCFA can reproduce the angular intensity of the semiclassical approach in Fig. \ref{Fig.compare} (a), indicating nearly perfect agreement between the two approaches. In contrast, the angle-unresolved LCFA performs poorly in describing the angular distribution, especially over azimuthal angle $\varphi$. The angle-resolved approaches predict a energy peak at $\theta_m=0.175$mrad, which is missing in the angle-unresolved case [Fig. \ref{Fig.compare} (d)], as well as   a spread of azimuthal angle over $\varphi=0$. In the case of angle-unresolved case, the photons are emitted along the electron momentum direction, which is in the $x-z$ plane for the linearly polarized planewave propagating along $z$. Therefore, the momentum of the photon has negligible $y$ component, i.e. $\varphi\approx0$. In contrast,
the difference in width of $dI/d\theta$ is negligible for different approaches. It is because the width is induced by the deflection of electron in the field $\theta_D\sim a_0/\gamma$, which is one order larger compared to  the correction of angular spread $\Delta\theta\sim 1/\gamma$.

\begin{figure}
    \includegraphics[width=0.5\textwidth]{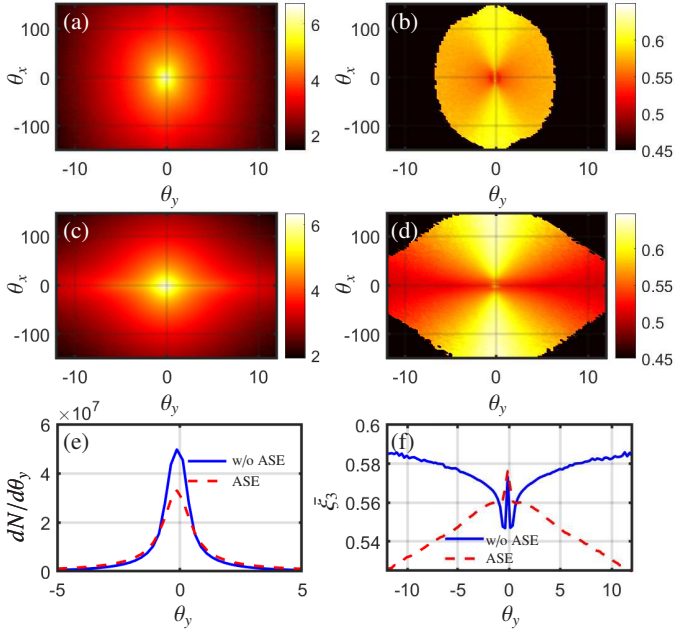}
    \begin{picture}(300,20)
     \put(23,240){\color{white}(a)}
    \put(60,175){$\theta_y$}
    \put(2,219){\rotatebox{90}{$\theta_x$}}
	\put(147,240){\color{white}(b)}
	\put(184,175){$\theta_y$}
    \put(126,219){\rotatebox{90}{$\theta_x$}}
    \put(23,161){\color{white}(c)}
    \put(60,96){$\theta_y$}
    \put(2,140){\rotatebox{90}{$\theta_x$}}
	\put(147,161){\color{white}(d)}
	\put(184,96){$\theta_y$}
    \put(126,140){\rotatebox{90}{$\theta_x$}}
    \put(23,81){(e)}
    \put(65,16){$\theta_y$}
    \put(2,46){\rotatebox{90}{$dN/d\theta_y$}}
    \put(87,75){\fontsize{5.5pt}{\baselineskip}\selectfont w/o ASE}
    \put(87,67){\fontsize{5.5pt}{\baselineskip}\selectfont  ASE}
    \put(147,81){(f)}
    \put(189,16){$\theta_y$}
    \put(118,58){\rotatebox{90}{$\bar{\xi_3}$}}
    \put(2,46){\rotatebox{90}{$dN/d\theta_y$}}
    \put(183,45){\fontsize{5.5pt}{\baselineskip}\selectfont w/o ASE}
    \put(183,38){\fontsize{5.5pt}{\baselineskip}\selectfont ASE}
	\end{picture}
    \caption{Angular distribution of photon density $\text{log}_{10}d^2N/d\theta_x/d\theta_y$ $ (\text{mrad}^{-2})$ (left column) and polarization $\xi_3$ (right column) vs $\theta_x$ and $\theta_y$: (top row) for angle-unresolved LCFA, and (middle row) for angle-resolved LCFA. The angular distribution of photon density $dN/d\theta_y$ $ (\text{mrad}^{-1})$ (e) and polarization $\xi_3$ (f) vs $\theta_y$ for angle-unresolved LCFA (blue solid line) and angle-resolved LCFA (red dashed line).}    
    \label{Fig.pho}
\end{figure}

\begin{figure}
    \includegraphics[width=0.5\textwidth]{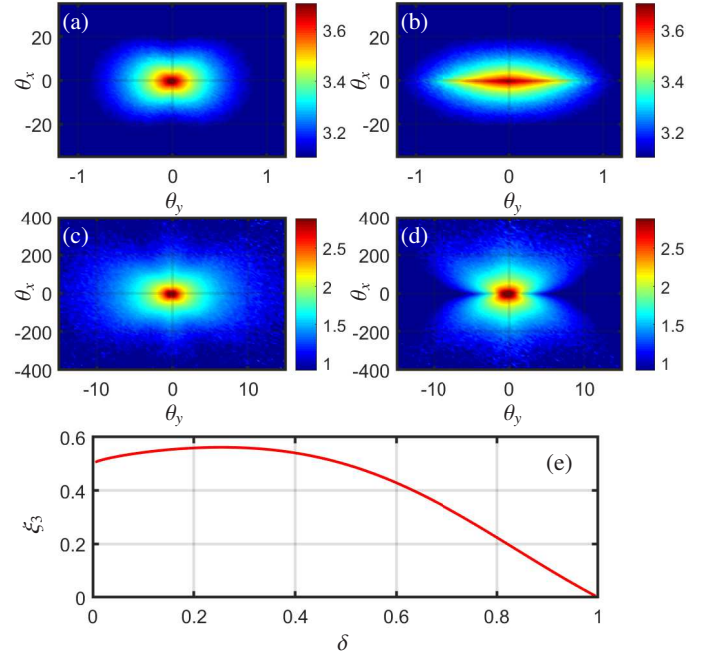}
    \begin{picture}(300,20)
   \put(19,248){\color{white}(a)}
    \put(58,180){$\theta_y$}
    \put(1,225){\rotatebox{90}{$\theta_x$}}
	\put(147,248){\color{white}(b)}
	\put(186,180){$\theta_y$}
    \put(128,225){\rotatebox{90}{$\theta_x$}}
    \put(19,167){\color{white}(c)}
    \put(58,100){$\theta_y$}
    \put(1,145){\rotatebox{90}{$\theta_x$}}
	\put(147,167){\color{white}(d)}
	\put(186,100){$\theta_y$}
    \put(128,145){\rotatebox{90}{$\theta_x$}}
    \put(202,81){(e)}
    \put(123,13){$\delta$}
    \put(4,57){\rotatebox{90}{$\xi_3$}}
	\end{picture}
    \caption{Angular distribution of average energy for electrons  $d^2\overline{\varepsilon}/d\theta_x/d\theta_y$ $ (\text{mrad}^{-2})$ (top row) and photons $d^2\overline{\omega}/d\theta_x/d\theta_y$ $ (\text{mrad}^{-2})$ (middle row) vs $\theta_x$ and $\theta_y$ for angle-unresolved LCFA (left column) and angle-resolved LCFA (right column). (e) The spectral stokes parameter $\xi_3$ vs emitted photon energy $\delta=\omega/\varepsilon_i$.}   
    \label{Fig.xi3}
\end{figure}

\subsection{The impact of ASE on density and polarization of out-going particles}

\subsubsection{Angle-resolved photon emissions}

Next, we proceed to investigate the effect of ASE in a more realistic scenario.  For investigating the polarization effects, intense lasers with asymmetry are needed to prevent the cancellation of spin effects in adjacent half-cycles.
Here, we use a tightly focused elliptically polarized laser with beam waist size $w_0=5\lambda_0$, laser intensity $a_0=100$ and ellipticity $\epsilon=|E_y|/|E_x|=0.03$. The electron beam consists of $N_e=6\times10^6$ electrons, with energy divergence $\Delta\varepsilon_i=0.06$, angular divergences $\Delta\theta_i=0.3$mrad and $\Delta\varphi_i=1$mrad. The rest parameters are same as in Fig. \ref{Fig.compare}.

The ASE induces a significant angular spread over $\theta_y$ but have negligible effects on the angular distribution over $\theta_x$ [Figs. \ref{Fig.pho} (a) and (c)]. The total angle of electron deflection in $x$ direction can be estimated as $\theta_x\sim a_0/\gamma$, orders larger than ASE $\Delta\theta\sim 1/\gamma$. While the deflection angle along $y$ axis is comparable with ASE since $\theta_y\sim \epsilon a_0/\gamma\sim O(1)/\gamma$. This angular spread is more significant for photons with $\theta_x\approx0$. It is because most of the photons are emitted by high-energy electrons with a small deflection angle ($\theta_x\approx0$), including a considerable amount of soft photons that are emitted at low-intensity region of laser pulse \cite{blackburn2020radiation}.
Since the significance of ASE is inversely proportional to emitted photon energy [Fig. \ref{Fig.delta_theta}], these soft photons are spread out dramatically when emission angle is resolved, resulting in a significant angular spread over $\theta_y$ around $\theta_x\approx0$. Correspondingly, the high-energy region of electrons is stretched out along $\theta_y$ around $\theta_x\approx0$ as a result of angle-resolved radiation reaction of soft-photons emissions [Figs. \ref{Fig.xi3} (a) (b)].
The relative error to angular distribution of photon density by neglecting ASE is $(\Delta \theta_y-\Delta \theta_y^{un})/\Delta \theta_y\approx 22\%$, with $\Delta \theta_y$ and $\Delta \theta_y^{un}$ being the full width at half maximum of angular distribution for angle-resolved and unresolved approaches, respectively [Fig. \ref{Fig.pho} (e)].

Regarding to the photon polarization, the changing law of $\xi_3$ with $\theta_y$ is opposite in large angle region for different approaches [Fig. \ref{Fig.pho} (f)]. $\xi_3$ increases with $\theta_y$ for angle-unresolved LCFA but decreases for angle-resolved LCFA.
The $\gamma$ photons emitted by the unpolarized electrons are in a mixed state with $\bm{\xi}=(0,0,\xi_3)$. The average polarization of the emitted photons can be estimated with
\begin{equation}
\xi_{3}=\frac{\textrm{K}_{\frac{2}{3}}\left(\frac{2u}{3\chi}\right)}{-\int_{\frac{2u}{3\chi}}\textrm{K}_{\frac{1}{3}}\left(x\right)dx+\frac{\varepsilon^{2}+\varepsilon'^{2}}{\varepsilon'\varepsilon}\textrm{K}_{\frac{2}{3}}\left(\frac{2u}{3\chi}\right)},
\end{equation}
where $u=\omega/\varepsilon'$. As shown in [Fig. \ref{Fig.xi3} (e)], $\xi_3$ is inversely proportional to photon energy $\delta=\omega/\varepsilon_i$ except for the soft photons located at $\delta\leq 0.25$. In the angularly unresolved case, $\theta_y\propto a_{y}/\gamma$ is inversely proportional to emitted photon energy $\omega=\chi\varepsilon_i\sim a_0\gamma^2$. The high-energy photons are well collimated in the beam center while the low-energy photons are emitted with a larger angle [Fig. \ref{Fig.xi3} (c)]. Therefore, the polarization $\xi_3$ increases with the increase of $\theta_y$ [Fig. \ref{Fig.pho} (f)]. While in the angle-resolved case, the angular spread of the soft photons around $\theta_x\approx0$ could result in a reduction of average photon energy at large $\theta_y$. For instance, the soft photons distributed at $\theta_y=0$  move towards $\theta_y=\Delta\theta_y$,  resulting in a decrease of average energy of photons at $\Delta\theta_y$. Since soft photons are mostly distributed around $\theta_x\approx0$, the angular distribution of photon energy is severely distorted at this region [ Fig. \ref{Fig.xi3} (d)]. When the photon energy decrease to $\delta\leq 0.25$,  $\xi_3$ decreases with the decrease of $\delta$  [ Fig. \ref{Fig.xi3} (e)], which is responsible for the  decrease of $\xi_3$ with the increase of  $|\theta_y|$ at $|\theta_y|>1.3$mrad [Fig. \ref{Fig.pho} (f)].

Note that, even though the ASE affects the angular distribution of photon density and polarization, the average polarization over all the emitted photons $\overline{\xi}_3=\overline{\xi}^{un}_3\approx0.56$ is unchanged. In our scheme, $\xi_3$ of a emitted photon is solely determined by its energy, which is irrelevant to ASE. The role of ASE is changing the momentum direction of a photon, leading to a variation of angular distribution of $\xi_3$. However, the average polarization of the photon beam is unaffected by ASE.

\begin{figure}
    \includegraphics[width=0.5\textwidth]{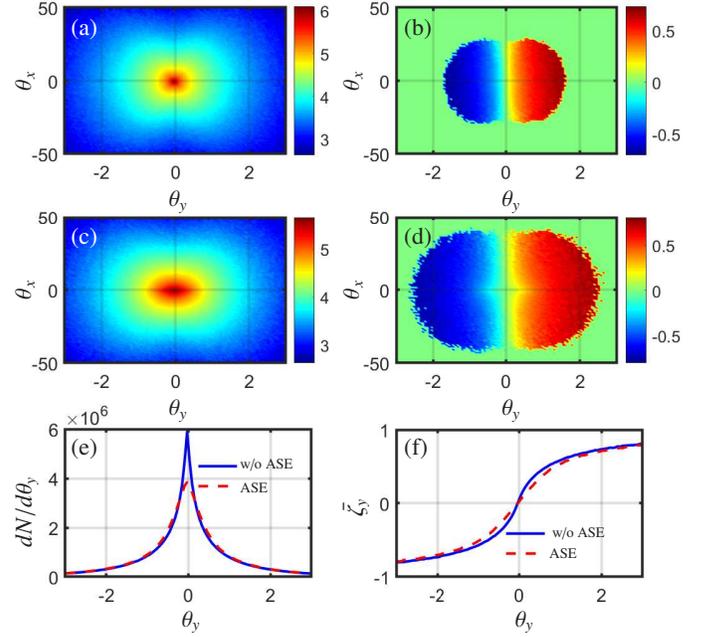}
    \begin{picture}(300,20)
  \put(23,240){\color{white}(a)}
    \put(60,175){$\theta_y$}
    \put(2,219){\rotatebox{90}{$\theta_x$}}
	\put(149,240){(b)}
	\put(186,175){$\theta_y$}
    \put(126,219){\rotatebox{90}{$\theta_x$}}
    \put(23,161){\color{white}(c)}
    \put(60,96){$\theta_y$}
    \put(2,140){\rotatebox{90}{$\theta_x$}}
	\put(149,161){(d)}
	\put(186,96){$\theta_y$}
    \put(126,140){\rotatebox{90}{$\theta_x$}}
    \put(23,81){(e)}
    \put(65,16){$\theta_y$}
    \put(2,46){\rotatebox{90}{$dN/d\theta_y$}}
    \put(87,76){\fontsize{5.5pt}{\baselineskip}\selectfont w/o ASE}
    \put(87,67){\fontsize{5.5pt}{\baselineskip}\selectfont ASE}
    \put(149,81){(f)}
    \put(191,16){$\theta_y$}
    \put(126,58){\rotatebox{90}{$\bar{\zeta_y}$}} 
    \put(205,50){\fontsize{5.5pt}{\baselineskip}\selectfont w/o ASE} 
    \put(204,42){\fontsize{5.5pt}{\baselineskip}\selectfont ASE} 
	\end{picture}
    \caption{Angular distribution of electron density $\text{log}_{10}d^2N/d\theta_x/d\theta_y$ $ (\text{mrad}^{-2})$ (left column) and polarization $\zeta_y$ (right column) vs $\theta_x$ and $\theta_y$:  (top row) for angle-unresolved LCFA, and (middle row) for angle-resolved LCFA. The angular distribution of electron density $dN/d\theta_y$ $ (\text{mrad}^{-1})$ (e) and polarization $\zeta_y$ (f) vs $\theta_y$: for angle-unresolved LCFA (blue solid line) and angle-resolved LCFA (red dashed line). }   
    \label{Fig.ele}
\end{figure}
\begin{figure}[t]
    \includegraphics[width=0.5\textwidth]{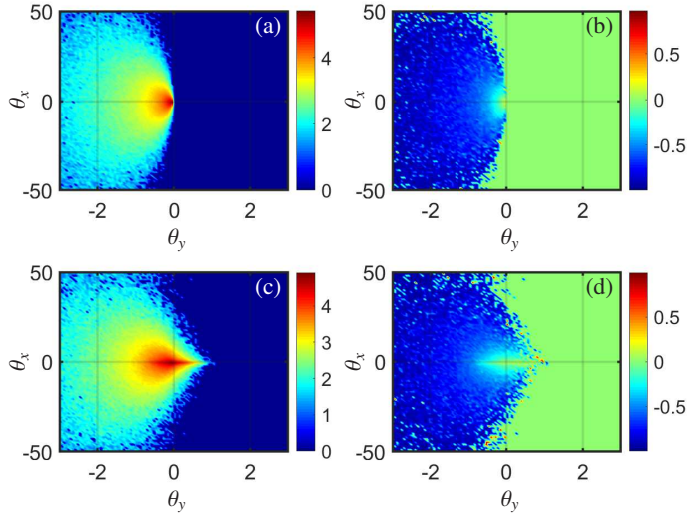}
    \begin{picture}(300,20)
    \put(90,204){\color{white}(a)}
    \put(57,124){$\theta_y$}
    \put(-3,175){\rotatebox{90}{$\theta_x$}}
	\put(215,204){(b)}
	\put(183,124){$\theta_y$}
    \put(124,175){\rotatebox{90}{$\theta_x$}}
    \put(90,105){\color{white}(c)}
    \put(57,25){$\theta_y$}
    \put(-3,76){\rotatebox{90}{$\theta_x$}}
	\put(215,105){(d)}
	\put(183,25){$\theta_y$}
    \put(124,76){\rotatebox{90}{$\theta_x$}}
	\end{picture}
    \caption{Angular distribution of electron density $\text{log}_{10}d^2N/d\theta_x/d\theta_y$ (mrad$^{-2}$)  (left column) and polarization $\zeta_y$ (right column) vs $\theta_x$ and $\theta_y$ with photon emissions at  $B_y>0$:  (upper row) for angle-unresolved LCFA, and  (bottom row) for angle-unresolved LCFA. }   
    \label{Fig.spread}
\end{figure}

\subsubsection{Electrons with angle-resolved radiation reaction}

The effects of ASE on electrons density and polarization are illustrated in Fig. \ref{Fig.ele}. The variation of angular distribution of electron density is similar with that of photons [Figs. \ref{Fig.ele} (a) (c) and (e)], but not for the polarization [Figs. \ref{Fig.ele} (b) (d) and (f)]. In both approaches, the electron beam is split along the propagation direction into two oppositely transversely polarized parts due to the spin-dependent radiation force \cite{li2019ultrarelativistic,dai2022photon,wan2020ultrarelativistic}. The electrons with $\zeta_y>0$ ($\zeta_y<0$) are more likely to emit photons at $B_y<0$ ($B_y>0$) due to asymmetric radiation probability. In an elliptically polarized laser field, the electron momentum $p_y$ and magnetic field $B_y$ is correlated as $p_y\cdot B_y<0$. Assuming the photons are emitted along the momentum direction of emitting electrons, the photons emitted at $B_y<0$ ($B_y>0$) could exert  a radiation kick to deflect the electrons towards $p_y<0$ ($p_y>0$), and finally the electron obtains $\zeta_y>0$ ($\zeta_y<0$) at $\theta_y>0$ ($\theta_y<0$). It is assumed to be a good approximation for ultrarelativistic electrons. However, the splitting angle of the electrons with opposite spin is in the order of milliradians, comparable with the emission angle $\Delta\theta\sim 1/\gamma$ for MeV electrons with $\gamma\leq10^3$. In the angle-resolved case, the polarization of electrons distributed at the beam center decreases. It is because the high energy region is stretched out around $\theta_x\approx0$ [Fig. \ref{Fig.xi3} (b)], and the polarization is  roughly inversely proportional to emitted photon energy, leading to a decrease of polarization around $\theta_x\approx0$. Meanwhile, the low-polarization gap between the splitting electrons is broadened. Since the correlation of polarization and radiation reaction force in an elliptically polarized laser field can be breakdown by angular spread for small angle electrons around $\theta_y\approx0$. For instance, the photons emitted at $B_y>0$ could exert a radiation kick to deflect electrons towards $p_y<0$ instead of $p_y>0$ due to the angular spread. Finally the distribution of electron with $\zeta_y<0$ is extended to $\theta_y>0$, averaging out the polarization of electrons polarized with $\zeta_y>0$ at $\theta_y>0$ [Fig. \ref{Fig.spread}].
The relative error to electron polarization induced by neglecting ASE is $(\zeta_y-\zeta_y^{un})/\zeta_y\approx 10.82\%$, with $\zeta_y$ and $\zeta_y^{un}$ being the average polarization for angle-resolved and unresolved approaches, respectively.

\section{conclusion}
We developed an angle-resolved Monte-Carlo method for investigating the impact of angular spread during nonlinear Compton scattering, by employing the angle-resolved radiation probabilities in the local constant field approximation. We assessed the validity of the commonly used strong-field QED simulation, which based on the angle-unresolved LCFA probability and the assumption that photons propagate along the momentum direction of the emitting electrons. In a linearly polarized planewave, the angular distribution of emitted photons obtained with angle-resolved LCFA approach agrees with the semiclassical approach without LCFA, while the angle-unresolved LCFA approach fails in describing the angular distribution along the magnetic field direction, where deflection angle $\theta_D\sim0$. For nonlinear Compton scattering in a realistic elliptically polarized laser field, the deflection angle induced by laser field along $y$ is rather small as $\theta_y\sim \epsilon a_0/\gamma$ with ellipticity $\epsilon\sim O(10^{-2})$. In this case, neglecting the emission angle between photons and electrons could result in a relative error of $\sim$55$\%$ in angular distribution of electrons.
Meanwhile, the angular spread at emissions could also affect angular distribution of polarization for the out-going particles. The soft photons at beam center are relocated towards large angle region, inducing a decrease of average photon energy at $|\theta_y|>1.3$mrad and consequently a decrease of photon polarization  at this region. The angular distribution of electron density and polarization are also affected by angular spread as a consequence of angle-resolved radiation reaction.  The high energy electrons are stretched to a larger $\theta_y$ around $\theta_x\approx0$ due to the angle spread of soft photon emissions, resulting in a decrease of polarization near the beam center. 
More importantly, since the polarization of electrons in elliptically polarized laser is generated by the spin-dependent radiation reaction, the angle spread could change the direction of radiation reaction force, breaking the correlation of electron spin and deflecting angle in an elliptically polarized laser field. This is a nontrivial effect for electrons around $\theta_y\approx0$, which causes a broaden of low-polarization gap between electron ensembles with opposite polarization.
Therefore, the ASE is crucial for accurate descriptions of angular distribution of density and polarization for out-going particles, and could impact the average polarization of the particle beams if the considered laser driven polarization scheme is sensitive to emission angles. \\

{\it Acknowledgement:}
This work is supported by the National Natural Science Foundation of China (Grants No. 12074262), the National Key R\&D Program of China (Grant No. 2021YFA1601700), the Shanghai Rising-Star Program, and the project Advanced
Research using High Intensity Laser Produced Photons and Particles (ADONIS) (Project No. CZ.02.1.01/0.0/0.0/16 019/0000789) from the European Regional Development Fund.

\bibliography{prp}

\end{document}